\documentclass[3p]{elsarticle}
\usepackage{amssymb,amsmath} 
\usepackage{graphicx} 
\usepackage{caption}

\usepackage{hyperref}
\hypersetup{
    colorlinks=true,
    linkcolor=red,
    filecolor=magenta,      
    urlcolor=cyan
}

\newcommand{\ex}[1]{ 
\mathbb{E}_{\mathbf{x}} \left( #1 \right)
}

\begin{document}
 
\title{A simple individual-based population growth \\ model with limited resources}

\author[1]{Luis R. T. Neves \corref{cor1}}
\ead{rodrigo2neves@gmail.com}
\author[1]{Leonardo Paulo Maia}

\cortext[cor1]{Corresponding author}
\address[1]{Instituto de F\'isica de S\~ao Carlos, Universidade de S\~ao Paulo, S\~ao Carlos, Brazil}

\begin{abstract}

	We address a novel approach for stochastic individual-based modelling of a single species population. Individuals are distinguished by their remaining lifetimes, which are regulated by the interplay between the inexorable running of time and the individual's nourishment history. A food-limited environment induces intraspecific competition and henceforth the carrying capacity of the medium may be finite, often emulating the qualitative features of logistic growth. Inherently non-logistic behaviour is also obtained by suitable change of the few parameters involved, composing a wide variety of dynamical features. Some analytical results are obtained. Beyond the rich phenomenology observed, we expect that possible modifications of our model may account for an even broader scope of collective population growth phenomena.

\end{abstract}


\maketitle

\section{Introduction}

Historically, population dynamics has acquired a much broader perspective along the development of mathematical modelling, playing a central role in the study of the spreading of epidemics, opinions, computer viruses and even fake news or characterizing the evolution of languages and cancer cell populations \cite{nowak-2006}. Notwithstanding the inherently nonlinear character of even the simplest interactions among individuals, a widespread practice when studying populations consists in building dynamical models by adding simple mathematical expressions thought of as ingredients, each corresponding to a basic populational behaviour. Roughly speaking, one expects the full model to keep the elementary ingredients but also being able to allow the emergence of more complex behaviours through nonlinearity. Aside the basic linear models for either growth or decay of a set of independent individuals, the simplest model for population dynamics is the logistic interaction among its members.

Within the scope of single-species modelling -- to which this manuscript is entirely dedicated --, there are two well-stablished deterministic logistic models, distinguished in the nature of their temporal evolution \cite{strogatz-2015,hirsch-etal-2012}. Verhulst's exactly solvable continuous time logistic equation, $\dot{x}=\lambda x (1-x/K)$, only presents monotonic trajectories of the total population $x(t)>0$ evolving from its initial size $x(0)$ towards the carrying capacity $K>0$. Far from this condition, it behaves roughly as an exponential, either decaying when $x(t)>>K$ or growing with rate $\lambda>0$ when $0<x(t)<<K$. The second deterministic paradigm, the logistic map, $x_{t+1}=\lambda x_t (1-x_t)$, is a discrete time dynamical system with a rich variety of behaviours regulated by the bifurcation parameter $\lambda > 0$. It can converge to a fixed point but its asymptotic behaviour can also consist of oscillations and it is even capable of achieving chaotic behaviour through a period-doubling route. Despite not being a proper discretization of the continuous-time version, such an extensive phenomenology rendered popularity \cite{may-1976} to the logistic map as an example of generation of complexity by simple models.

Regardless of their profound conceptual, mathematical and phenomenological distinctions, the Verhulst equation and the logistic map share at least two important, general features that may put them in the same class if we are to characterize population dynamics models. First, the only relevant variable concerning the state of the population is its size (or density). In other words, the population is completely described by the number of living individuals and, as a consequence, the dynamics of its temporal evolution can in no way depend on its internal structure or distinction among individuals. It is therefore assumed that the totality of the complex interactions amongst individuals and between population and environment may be encoded just in how the employed model maps the population size into its temporal evolution, in a kind of ``mean-field'' fashion. Following the literature, we shall refer to any model of this kind as a population-level (or population-based) model.  

A second, more obvious aspect is that both the logistic ODE and logistic map are deterministic dynamical systems: specifying the state of the system in a given instant of time suffices, at least in principle, to predict the exact outcome of an arbitrary time evolution.

Other notorious, although more specific models that fit within the same class are the Monod and Droop models for microbial growth in laboratory conditions \cite{wade-2016, shoener-2019}. These models may include other variables than the population size, such as nutrient concentration in the medium; however, when it comes to the description of the population itself, only its density is taken into account.

Deterministic, population-based models are of acknowledged historical and practical importance.  Nevertheless, since the 90's a notorious growth has been reported in the employment of a radically different approach which we may generically name individual-based modelling \cite{mckane-2004, deangelis-2005}. In this framework, rather than writing down some equation determining the time evolution of the population size as a function of it (and possibly some population-external variables), the starting point is the description of the elementary units of the system, i. e., individuals. The modeller then designs some idealized behaviour rule which governs the actions of this single individual, and expects the global system to follow some ecologically reasonable law. Besides, the mentioned individual-level behaviour rules are typically of probabilistic nature. Such rules may try to capture real actions performed by individuals, such as feeding, replicating, dying, or moving. We shall label any of these models an individual-based model (IBM). We ought to make it clear that many authors distinguish between individual-based and agent-based models \cite{mckane-2004}; we choose to ignore such distinction and just employ the term IBM to mean that the fundamental dynamical laws are formulated at the level of the individual, rather than that of the whole population. 

The fact that IBMs allow us to unveil ``macroscopic'' collective behaviour as emerging from ``microscopic'' individual one (and, sure enough, from the intricate network of interactions that takes place) is itself a tremendous advantage from the theoretical point of view. It provides, for instance, a natural framework to account for populations structured in space \cite{law-2003}; besides, there is the obvious fact that real populations do exhibit observable variations from one individual to another, which cannot be captured by population-level models \cite{mckane-2004, natalie-2017}. This approach has also shown great practical importance, providing much more accurate predictions on real ecological situations, also opening up the possibility of learning about the behaviour of individuals from observational data concerning the dynamics of the entire population \cite{deangelis-2005}. 

In this work, we do not discuss models which consider the spatial distribution of individuals; that is, no notion of space is explicitly taken into account. We shall refer to such restrict models as \textit{population growth} models, while the more general term \textit{population dynamics} might include spatial issues.

Some of the so-called logistic IBMs are designed in such a way as to recover the logistic growth curve in average, or in some deterministic limiting case; this, however, is often achieved through the imposition of individual behaviour rules that somehow mimic the mathematical structure of Verhulst's equation -- for instance, the probability of an individual replicating in a short interval of time being proportional to $1 - N/K$, where $N$ is the size of the population \cite{natalie-2017, nasell-2001, otso-2010}. Though useful and convenient, this approach is artificial in the sense that such probabilistic law is largely an \textit{ad hoc} construction, and the very meaning of the parameter $K$, for example, is clear only as long as the population-level model that gave birth to this individual-based one is known \textit{a priori}. In other words, it is difficult to make sense of such a law in the individual level by itself. 

So motivated, one of the goals of this work is to introduce an IBM based upon a simple, intuitively-justified set of individual behaviour rules which might reproduce a logistic-like curve as a genuinely emergent phenomenon. The same model, nonetheless, will give rise to many other different qualitative dynamical signatures, such as damped oscillation towards equilibrium, and even perpetual growth. The model is based upon the description of individuals that can reproduce asexually and indirectly compete for nourishment resources. We model a resource-limited environment, and the available food is distributed randomly among individuals; the finiteness of this food is what induces the referred competition effect and, consequently, may stop the vegetative growth of the population. Only three fixed parameters rule individual behaviour and, thence, determine the different collective growth patterns that emerge. We have been able to derive few analytical results, from which we highlight the equation for the carrying capacity of the medium as a function of the referred individual-level parameters. Due to the complexity of the system, however, much of the carried out characterization of the dynamics was based in the outcome of numerical simulations.

We now outline the structure of this manuscript. In \S \ref{sec:themodel} we introduce our model, defining precisely the abstract representation of individuals and the laws that rule their behaviour, immediately proceeding to the formal description adopted and the fundamental stochastic equation of the dynamics. \S \ref{sec:results} is the description of all our results, both analytical and numerical; they are subdivided in considerations about equilibrium (\S \ref{sec:equilibrium}), a deterministic analogue of the dynamics (\S \ref{subsec:det}), critical and supercritical regimes (\S \ref{subsec:crits}) and the effect of initial conditions (\S \ref{subsec:initcond}). Then, we discuss our results, in perspective with the possible connection to real-life populations and in the context of other works on population modelling (\S \ref{sec:disc}), finally summarizing our conclusions (\S \ref{sec:conc}).

\section{The model \label{sec:themodel}}

\subsection{Definition}

We start with a qualitative overview of the individual behaviour rules. We study a population of individuals evolving in discrete time, always subjected to an \textbf{aging} process and eventually benefited from \textbf{feeding} and performing \textbf{reproduction}.

At any time step, each individual is characterized by a single, \textit{dynamical} attribute: its \textit{current} remaining lifespan $\tau$, modelled as a positive integer in the set $\{ 1, ...,\omega \}$ and that, in the absence of feeding (by which an individual counteracts aging -- ``food'' means additional lifetime in this model), will certainly decrease one unit at each time step. The population size at time $t$, $N(t)$, may decrease from one time step to the next because some individuals may exhaust their remaining lifespans and die, but also may increase by reproduction events. 

Feeding is the only mechanism that may counteract senescence, but its distribution depends on population size -- in a way such as to capture the limitedness of environmental resources --  and will mimic competition in large populations. Across any unit time step, there is a maximum amount of food available for the whole population, $\alpha$ (a constant positive integer), and each individual will get either one unit of food or nothing, with \textit{dynamical} probabilities, dependent on population size. If $N(t) \leqslant \alpha$, everyone will get fed (there is no competition) and the unused excess $\alpha - N(t)$ plays no role in the dynamics. In case of overcrowding, individuals compete for food with equal strengths, regardless their ages. In general, the marginal probability of an individual receiving a unit of food is $\min{\left\{ 1,\alpha/N(t)\right\} }$.

Finally, takes place asexual reproduction, where each individual, no matter how old, originates a single newborn with probability $\rho$. Newborns enter the population at their prime, each one with remaining lifespan of $\omega$ units.

Thus, $\alpha, \omega \in \mathbb{N}$ and $\rho \in \left( 0,1\right)$ are our fixed parameters. Formally, at a given instant $t$ our population is a list $\left\lbrace \tau_j \right\rbrace_{j=1,...,N(t)}$ of remaining lifetimes for some arbitrary indexing of the population, whose time evolution is defined as the sequence of the following three operations:

    \begin{enumerate}
        \item \textbf{Feeding:} if $\alpha \geqslant N(t)$, we make $\tau_j \leftarrow \tau_j + 1, \forall j$. Otherwise, $\alpha$ distinct indices in $\left\{ 1,...,N(t) \right\}$ are drawn at random, and the update $\tau_j \leftarrow \tau_j + 1$ takes place for each drawn index (the remainder of the population is kept untouched at this step). Alternatively, for any $N(t)$, in a random order, individuals have a single shot at getting a unit of food, with probability $\min{ \left\{ 1,\alpha/N(t) \right\} }$, until either all getting fed or exhausting the capacity $\alpha$;
        \item \textbf{Aging:} for every $j$, $\tau_j \leftarrow \tau_j - 1$. Individuals fed in the previous step will have prevented ageing. If $\tau$ becomes zero for some individual, it dies and its entry is simply deleted from the list;
        \item \textbf{Reproduction:} any individual, with probability $\rho$, gives rise to a single descendant, born with remaining lifespan $\omega$.
    \end{enumerate}

By convention, the three manipulations above constitute \textit{one} time step, and, in general, we describe the state of the population \textit{between} time steps. Note also that, no matter how convenient have been an indexation of each individual in describing the above steps, it becomes clumsy trying to keep track of it across generations of a size-variable population. We will see that is much simpler adopting a coarse-grained description of the population, clustering individuals in $\omega$ classes, by age, with no loss of information relevant to the dynamics of $N(t)$.

\subsection{Formal description}

Although so defined, describing our population as a list of $\tau_i$'s is unnecessarily complicated (for instance, the very ``shape'' of the list, $N(t)$, is itself a dynamical variable). For, steps $1$ and $2$ above assure that a given individual's $\tau$ value never increases throughout a time step -- it either remains constant or decreases in a unit --; besides, individuals are always born at $\tau = \omega$. Thus, if we assure an initial condition such that $\tau_i \leqslant \omega$, $\forall i$, which we shall always do, it is certain that such constraint will remain true at any instant of time. Now we are allowed to partition our population with respect to the lifespans, writing down a list

    \begin{eqnarray}
        \mathbf{X}_t = \left(X^1_t, ..., X^\omega_t \right),
    \end{eqnarray}
where $X^j_t$ is the number of individuals for which $\tau = j$ at time $t$. The obvious advantage of this description is that now the dimension of the list (which specifies the state) is a fixed parameter, namely $\omega$. Our main goal is then to describe the mathematical features of the dynamics induced by steps 1{-}3 on the state $\mathbf{X}_t$. 

A straightforward translation of the individual-based dynamics into the terms of this class-structured description leads to the equations below, which encode all the information about the time evolution of $\mathbf{X}_t$:

\begin{eqnarray}
    \label{eq:class_std_dynamics}
	X^i_{t+1} = \begin{cases}
    (X^{i+1}_{t} - Q^{i+1}_{t}) + Q^i_{t}, & i = 1,..., \omega - 1; \\
     Q^i_{t} + R_{t}, & i = \omega, \\
    \end{cases}
\end{eqnarray}

where $Q_t^i$ is the number of fed $i-$th class individuals in the step $t \rightarrow t + 1$, and $R_t$ the total of newborns generated in the same step. We shall always denote random variables by capital Roman letters and fixed parameters by lowercase Greek letters. 

\section{Results \label{sec:results}}

\subsection{\label{sec:equilibrium}Equilibrium. Subcritical growth}

The stochastic dynamics is extremely simple so long as $N_t \equiv \sum_i X^i_t \leqslant \alpha$, for in such case Eqs. \eqref{eq:class_std_dynamics} reduce to $X_{t+1}^i = X_t^i, \, i < \omega$ and $X_{t+1}^\omega = X_t^\omega + R_t$; besides, $R_t$ is binomially distributed with parameters $(N_t, \rho)$. Thus $N_t$ grows exponentially in average and no steady state could be reached. We shall call this \textit{abundance regime}, in opposition to the \textit{competition regime} ($N_t > \alpha$); the latter exhibits a much richer dynamics, with which we shall be mainly concerned.
    
From Eqs. \eqref{eq:class_std_dynamics} it is easy to derive the expectation value of $\mathbf{X}_{t+1}$, \textit{conditioned} to $\mathbf{X}_t = \mathbf{x} \equiv (x^1,...,x^\omega)$ (i. e., for fixed $\mathbf{X}_t$). We can show (Appendix \ref{sec:trans_ex}) that, under competition regime,

    \begin{eqnarray}
        \label{eq:expectation}
	    \ex{ X^i_{t+1} } = \begin{cases}
       (\alpha/n)x^i + (1 -\alpha/n)x^{i+1}, & i < \omega; \\
         (\alpha/n)x^i + \rho \left[n - (1 - \alpha/n)x^1 \right], & i = \omega, \\
        \end{cases}
    \end{eqnarray}
where the subscript $\mathbf{x}$ stands for the referred conditioning.

We shall adopt a simple definition of stochastic equilibrium, in that a state $\mathbf{x}^*$ is said to be of equilibrium if it satisfies

    \begin{eqnarray}
        \mathbb{E}_{\mathbf{x}^*}\left( \mathbf{X}_{t+1} \right) = \mathbf{x}^*,
    \end{eqnarray}
that is, the condition $\mathbf{X}_t = \mathbf{x}^*$ is stationary \textit{in average}. Combining condition above with Eq. \eqref{eq:expectation} yields

    \begin{eqnarray}
        \label{eq:equilibrium}
        \mathbf{x}^* = \left( n^*/\omega,...,n^*/\omega \right),
    \end{eqnarray}
where

    \begin{eqnarray}
        \label{eq:eq_size}
        n^* \equiv \frac{(1+\rho)\alpha}{1-(\omega -1)\rho}
    \end{eqnarray}
is the equilibrium size of the population, analogous to the carrying capacity in the logistic model. In particular, Eq. \eqref{eq:equilibrium} shows that, in equilibrium, all ``classess'' are equally populated. It is immediately noted that Eq. \eqref{eq:eq_size} is potentially pathological in that, if $1 - (\omega - 1)\rho \leqslant 0$, it could not possibly represent the size of a population. Indeed, the very definition of our population (along with its individual behaviour rules) cannot make sense of such thing as negative population size; thus, what Eq. \eqref{eq:eq_size} suggests is that, for a certain regime of parameters, no steady state can hold whatsoever. For that reason we shall call conditions $1 - (\omega-1)\rho > 0$, $= 0 $ and $< 0$ \textit{subcritical}, \textit{critical} and \textit{supercritical} regimes, respectively. 

\begin{figure}
\centering
\includegraphics[width=0.45\linewidth]{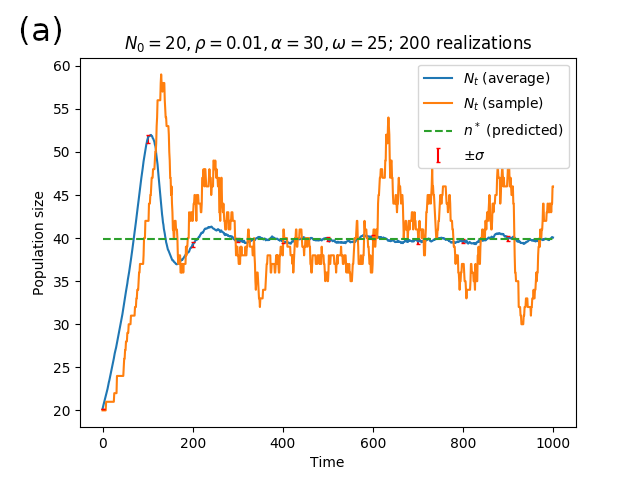} \includegraphics[width=0.45\linewidth]{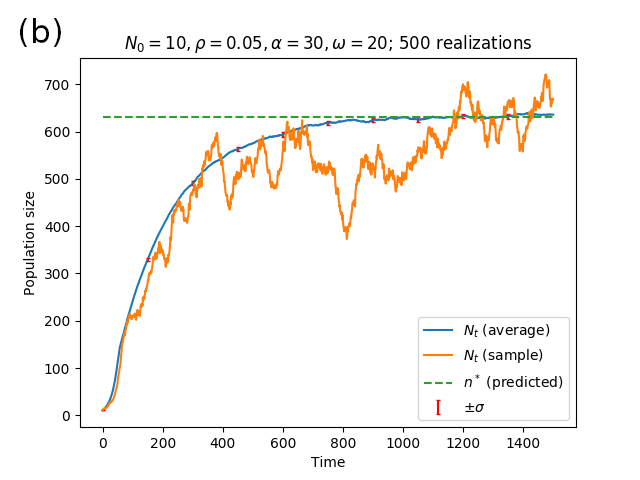}
\caption{Simulation results in subcritical regime.}\label{fig:1}
\end{figure}

Analyzing the subcritical case as our starting point, we show that numerical simulations reinforce our analytical predictions. (Initial condition is always $\mathbf{X}_0 = (0, ..., 0, N_0)$, unless otherwise stated.) In Fig. \ref{fig:1}, typical behaviour of $N_t$ in subcritical regime is depicted, for two different parameter choices. Population size indeed reaches the steady-state value given by Eq. \ref{eq:eq_size}. 

As it turns out, oscillations in population size are not only of stochastic nature (as seen in a single realization), rather surviving to the averaging of many simulations, as one sees in Fig. \ref{fig:1} (a). Such oscillations, however, do not always take place, even in the subcritical case; Fig. \ref{fig:1} (b) shows instead a monotonous growth toward equilibrium. Actually, it was verified that the occurrence of oscillatory behaviour may depend even upon the initial conditions; we shall return to this point in \S \ref{subsec:initcond}.

\begin{figure}		
\centering
\includegraphics[width=0.6\linewidth]{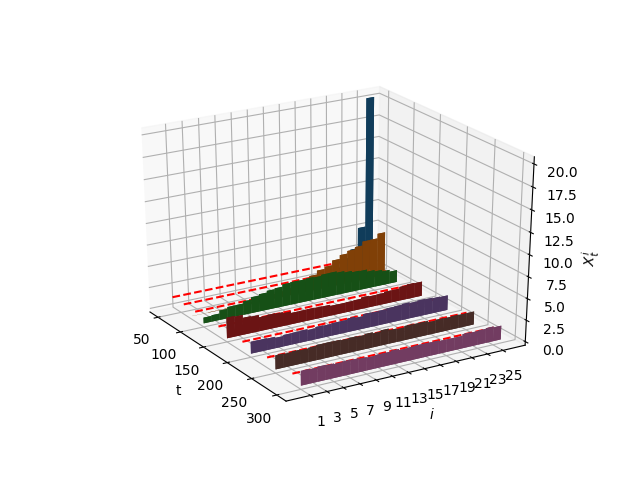} 
\caption{\label{fig:2}Evolution of $\mathbf{X}_t$ for the same data set of Fig. \ref{fig:1} (a). Red, dashed lines assign the predicted equilibrium value of $X^i$, namely $n^*/\omega$.}   
\end{figure}

Of course, the evolution of $N_t$ alone does not tell much about the dynamics of $\mathbf{X}_t$, which has to be visualized at any given instant as a histogram. Fig. \ref{fig:2} shows, for the same dataset of Fig. \ref{fig:1} (a), the state $\mathbf{X}_t$ for some values of $t$. It is shown that equilibrium state given by Eqs. \eqref{eq:equilibrium} and \eqref{eq:eq_size} is indeed reached for large values of $t$.

\subsection{Deterministic map \label{subsec:det}}

\begin{figure}
\centering
\includegraphics[width=0.45\linewidth]{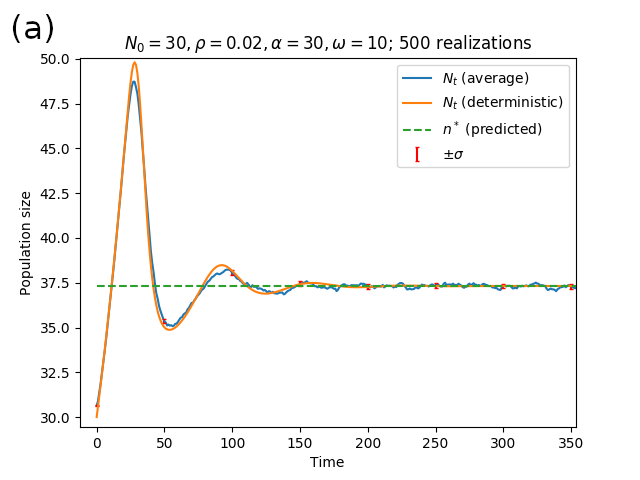} \includegraphics[width=0.45\linewidth]{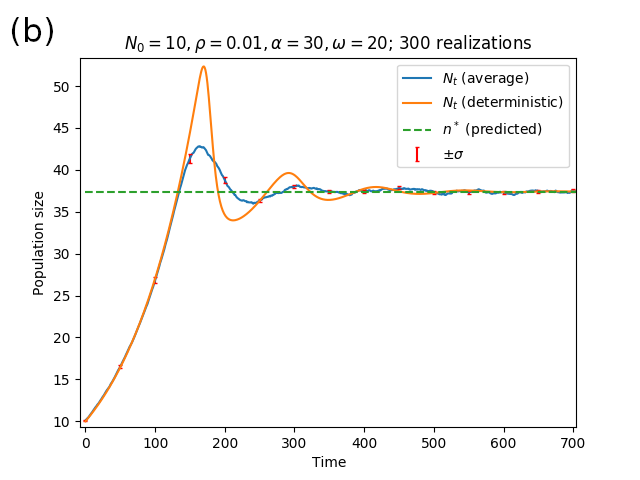}

\caption{\label{fig:3}Comparison between average of stochastic dynamics and deterministic map, in both a case where transient fluctuations are in good agreement and another where they are not.}
\end{figure}

Inspired by Eq. \eqref{eq:expectation}, one may define a deterministic map that partially captures the features of our original, stochastic system. Specifically, define a state $\mathbf{y}_t = (y^1_t, ..., y^\omega_t)$ which evolves according to the law

    \begin{eqnarray}
        \label{eq:deterministic}
	     y^i_{t+1} = \begin{cases}
       (\alpha/n_t)y^i_t  + (1 -\alpha/n_t)y_t^{i+1}, & i < \omega; \\
         (\alpha/n_t)y_t^i + \rho \left[n_t - (1 - \alpha/n_t)y_t^1 \right], & i = \omega, \\
        \end{cases}
    \end{eqnarray}
if $n_t \equiv \sum_i y^i_t > \alpha$, and $y^i_{t+1} = y^i_t + \delta_{i,\omega}\rho n_t$ otherwise. Such dynamics is by construction identical to the averaged result of that of Eqs. \eqref{eq:class_std_dynamics} only in abundance regime or in steady state; how close the two are in any other dynamical condition is, \textit{a priori}, a question to be answered numerically, given our lack of stronger analytical results for both systems. It turns out that the deterministic map nicely emulates the stochastic one for some choices of the parameters, while, for others, the transient fluctuations widely differ in amplitude (see Fig. \ref{fig:3}).

\subsection{Critical and supercritical regimes \label{subsec:crits}}

\begin{figure}
\centering
\includegraphics[width=0.5\linewidth]{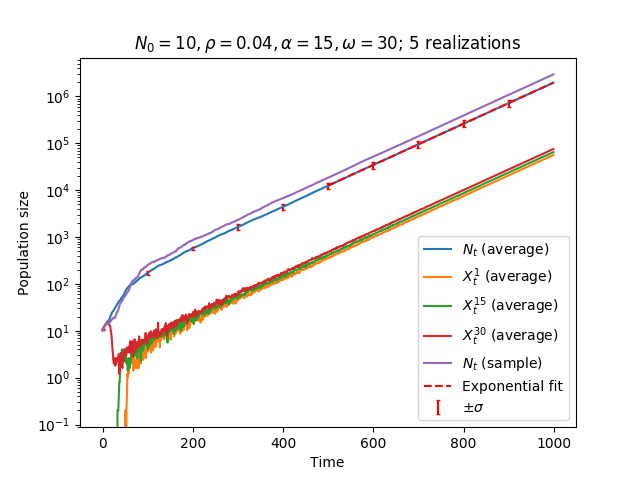}

\caption{\label{fig:4}Simulation results for a supercritical choice of parameters. Exponential fit was obtained by linear regression on $\log{\left \langle N_t \right \rangle}$, with time ranges previously chosen so as to disregard transient growth. The obtained angular coefficient was $a = 1.013 \times 10^{-2}$ with determination coefficient $R^2 = 0.999997$, justifying the claim that supercritical growth is exponential.}
\end{figure}

As seen in \S \ref{sec:equilibrium}, no steady state may be reached if $(\omega - 1)\rho = 1$ or $(\omega - 1)\rho > 1$. Simulations revealed that in the latter case (supercritical) a transient growth is followed by steady, exponential growth, as illustrated in Fig. \ref{fig:4}. Other choices of parameters in the same region have always exhibited the same behaviour.

\begin{figure}
\centering
\includegraphics[width=0.5\linewidth]{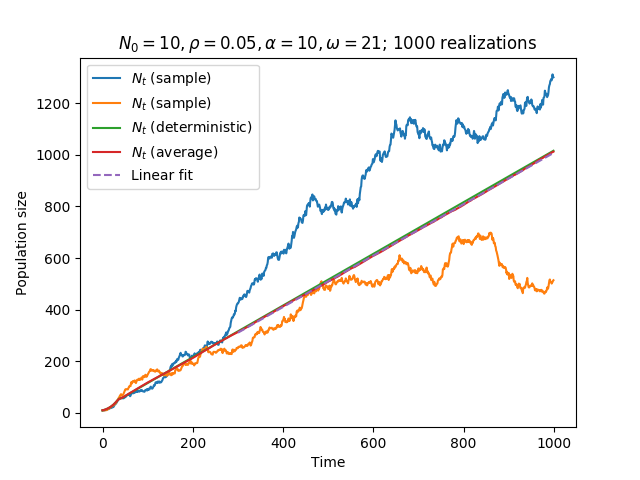}

\caption{\label{fig:5}Simulation results for population growth in critical case. The three straight lines (simulation average, deterministic map and adjusted line) are virtually undistinguishable. The slope obtained by linear regression was $0.995$ with determination coefficient $R^2 = 0.999$, in large agreement with the predicted value $2 \rho \alpha = 1$. Error bars have been omitted for being essentially of the width of the lines.}
\end{figure}

\begin{figure}
\centering
\includegraphics[width=0.45\linewidth]{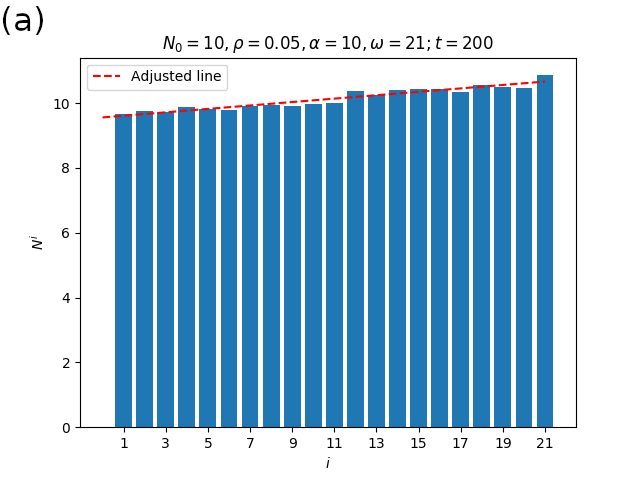}\includegraphics[width=0.45\linewidth]{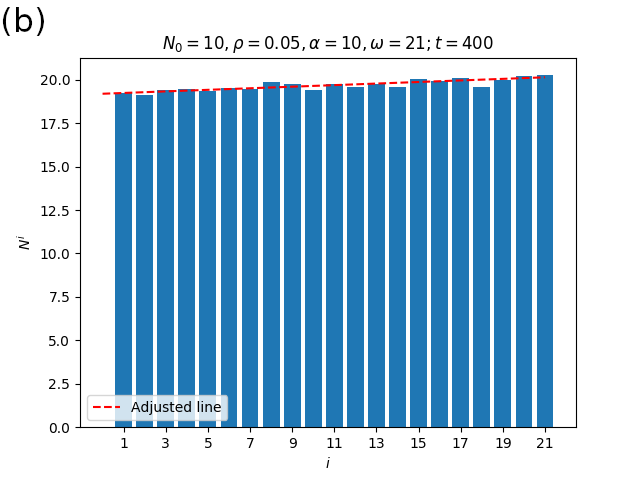}
\includegraphics[width=0.45\linewidth]{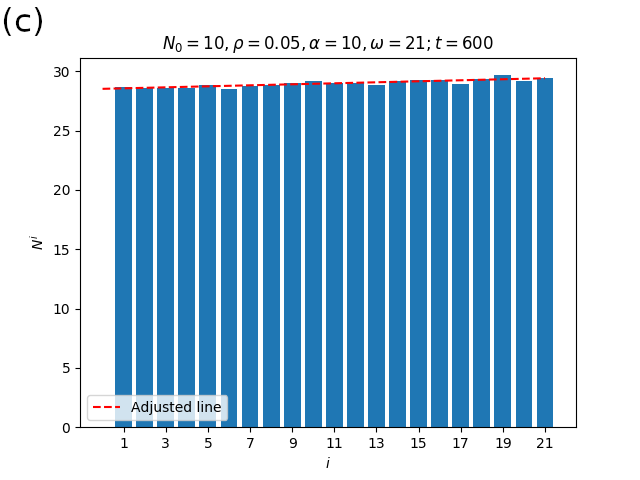}\includegraphics[width=0.45\linewidth]{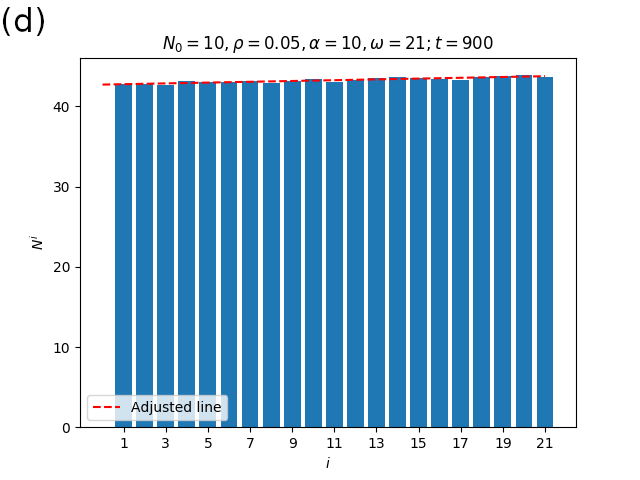}

\caption{\label{fig:6}Snapshots of histograms representing the average state $\mathbf{X}_t$ for the same simulation data set of Fig. \ref{fig:5} at different times. In each case, a straight line was fitted by linear regression. The numerical outputs were: (a) $a = 5.27 \times 10^{-2}, R^2 = 0.906$; (b) $a = 4.51 \times 10^{-2}, R^2 = 0.731$; (c) $a = 4.25 \times 10^{-2}, R^2=0.736$; (d) $a = 4.99\times 10^{-2}, R^2 = 0.786$. The predicted asymptotic value of the slope is $a = 2\rho\alpha/\omega \approx 4.76 \times 10^{-2}$. Not shown, deterministic map exhibits an even finer adjustment to the expected slope.} 
\end{figure}

In its turn, critical parameter regime exhibits linear growth following the transient. By means of an \textit{ansatz} motivated by such numerical observation, we have successfully calculated the asymptotic time dependence of deterministic $\mathbf{y_t}$ in this regime, namely 

    \begin{eqnarray}
        y^i_t = 2\frac{\rho \alpha}{\omega}t + 2\frac{\rho \alpha}{\omega}i + b^0,
    \end{eqnarray}
whence the population size $n_t = \sum_i y^i_t$ follows

    \begin{eqnarray}
        n_t = 2\rho \alpha t + \rho \alpha (\omega + 1) + \omega b^0,
    \end{eqnarray}
$b^0$ being an undetermined constant (see Appendix \ref{sec:critical} for proof). Simulations confirm our predictions, besides showing that the behaviour of $\mathbb{E}(\mathbf{X_t})$ is finely captured by the deterministic map in critical case (see Figs. \ref{fig:5} and \ref{fig:6}).

\subsection{Effect of initial conditions \label{subsec:initcond}} 

\begin{figure}
\centering
\includegraphics[width=0.45\linewidth]{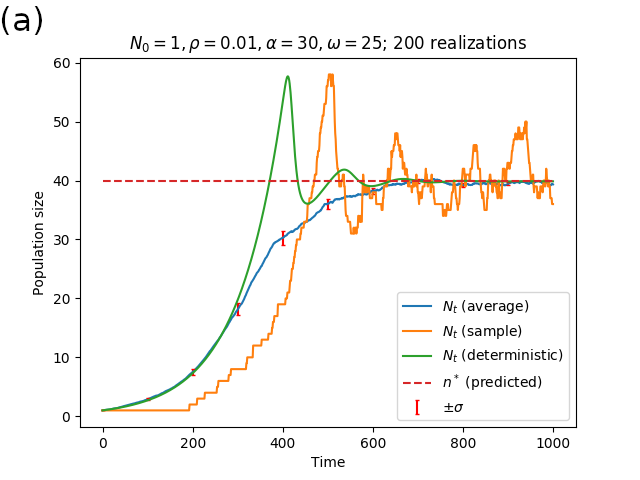}
\includegraphics[width=0.45\linewidth]{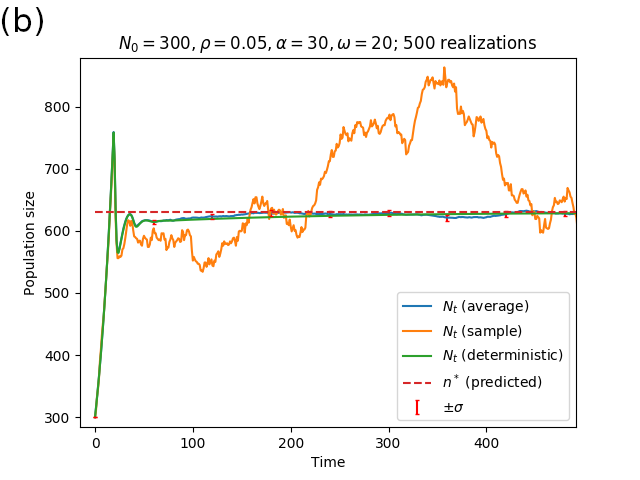}
\caption{\label{fig:7}Simulation results for same choices of parameters as in Fig. \ref{fig:1}, but different initial sizes, inducing qualitatively diverse transient dynamics (as to the presence or absence of damped oscillations).}
\end{figure}

We have verified previously (\S \ref{sec:equilibrium}) that transient oscillations may or may not occur in the subcritical regime, depending on the particular choice of parameters. As indicated, however, initial conditions may also play a determinant role in this feature of the dynamics. In Fig. \ref{fig:7} we show that, by suitably changing the initial population size ($N_0$) alone, we can control such property for both sets of parameters simulated in Fig. \ref{fig:1}. Fig. \ref{fig:7} (a) also shows that these oscillations may be present in the deterministic map even when the averaging of many realizations of the stochastic model shows monotonous growth; Fig. \ref{fig:7} (b), on the other hand, is an example of good agreement between both curves. 

As we are not concerned with the characterization of transient growth in critical and supercritical regimes, there is nothing to be discussed about the role of initial conditions in those cases.

\section{Discussion \label{sec:disc}}

In \S \ref{sec:results} we have shown that a myriad of qualitatively different patterns of population size growth may emerge from our individual-based model. For instance, the damped oscillations often shown in the subcritical regime (Figs. \ref{fig:1} (a), \ref{fig:3} and \ref{fig:7} (b)) resemble those that experimental microbiologists have observed decades ago for microorganisms growing in chemostats \cite{maas-1978}. Unexplained by the ``classical'' models of Verhulst and Monod, such oscillations have later been modelled with systems of differential equations, in an approach that explains them by taking into account an adaptive mechanism of individual cells as a response to varying nourishment conditions \cite{tang-1997}. Our model provides an alternative explanation for the same phenomenon; specifically, we may point at least two conditions that were necessary to promote damped oscillations, as in opposition to simple, prototypical models. One of them is contained in the very definition of our idealized system: the internal structure of the population, accounting for the biologically obvious fact that different individuals may live for shorter or longer times depending on how successful they are in the struggle for resources. Secondly, we have shown (\S \ref{subsec:initcond}) that, even for suitably chosen dynamical parameters, the occurrence of damped oscillations depends on ``special'' initial conditions. This might correspond, in a real ecological situation, to a population that emigrated from a region of abundant food resources to another of limited ones.

Of much simpler nature and interpretation is the growth pattern depicted in Figs. \ref{fig:1} (b) and \ref{fig:7} (a), where early exponential growth is followed by an inversion of the concavity and a steady approach to the carrying capacity. This is in qualitative correspondence to the paradigmatic logistic growth, broadly endorsed by empirical evidence concerning populations of species as diverse as flies \cite{krebs-2014}, protozoans \cite{krebs-2014}, and humans \cite{pearl-1920}. Our model, thus, may be regarded as capable of making logistic-like growth emerge from simple, biologically reasonable individual-level behaviour rules. It is noteworthy that many of the so-called individual-based logistic models achieve the desired population-level behaviour only through the arbitrary imposition of less natural rules, such as birth and death probabilities depending on population density in a carefully chosen fashion that is indeed reminiscent of the logistic equation. Examples of works that include this approach in more or less central positions are Refs. \cite{natalie-2017, nasell-2001, otso-2010}. 

Still in the case of subcritical regime, one important signature of our model is the expression for $n^*$, the carrying capacity of the environment, in terms of the fixed parameters. Within this region of the parameter space, $n^*$ is an increasing function of  $\rho, \alpha$ and $\omega$, as expected. Moreover, the dependency on $\alpha$, the only easily-controlled parameter in a hypothetical laboratory situation, is as simple as it could be. 

We now attempt to discuss and interpret the intriguing case of supercritical growth. Sure enough, one should be careful in trying to make sense of such thing as unlimited, exponential growth under resource-limited environmental conditions. Strictly speaking, this is undoubtedly a signature of a rather artificial feature of our model, namely the fact that individuals are always born with a predetermined (in this case, presumably long) ``life expectancy'', despite of their parent's. Indeed, we are forced to acknowledge that in a sense this is a border of the scope of validity of the model. However, we can reinterpret the supercritical growth as being meaningful (i. e., still eligible to describe some real phenomenon) only for a finite range of time, much like the Malthusian vegetative growth model $\dot{y} = \rho y$ may be valid only as long as the population size $y$ is small enough. 

A natural objection arises: is our model, in the supercritical regime, merely reproducing the same infamous perpetual exponential growth just in an infinitely more complicated manner than the oldest and simplest mathematical model of population dynamics does? Not quite, as we shall argue. First, we recall that early exponential growth, conditioned to the abundance of food, is always verified in our model, regardless of the parameter regime, as we have verified even analytically in \S \ref{sec:equilibrium}. The fact that \textit{another} period of exponential growth may be achieved, even away from the abundance regime (understood in the precise sense of \S \ref{sec:equilibrium}), is actually an indication that a different phenomenon is taking place in the dynamics of the internal structure of our system. Whilst the abundance regime is a situation in which no individual is deprived of food, whence none of them may possibly die -- and, as a consequence, they all keep replicating in a vegetative pace --, what happens in supercritical growth \textit{has} to be radically different. For, in such setting, the size of the population may be orders of magnitude larger than the available food per time step (see Fig. \ref{fig:4}); clearly, for large enough values of $N$, a typical individual will never get fed at all in its lifetime. Such individual, then, will die after a time interval of length $\omega$, generating an average of $\rho \omega$ offsprings in the meantime. Thus, the net contribution of this tracked individual to the population size, $\omega$ time steps after its birth, is $\rho \omega - 1$; the supercritical condition $\rho(\omega-1) > 1$ implies $\rho \omega - 1 > \rho > 0$, elucidating why, despite the virtual absence of food, this combination of large birth rate and life span assures a sustained growth of the population. Of course, unveiling such a mechanism is in no way a validation of the model in the referred regime; it remains an open task to find a real-life situation that might be modelled by it. 

Finally, critical growth is the most delicate case. It has the same undesirable feature of indefinite growth as does the supercritical regime; in qualitative terms, what differs one from the other is that, in the supercritical case, the dependence of $\left \langle N_t \right \rangle$ on $t$ appears linear only for a limited lapse of time, subsequently revealing its higher order corrections, only adequately described by an exponential function. In the critical case, for its turn, the time for those corrections to become evident may be regarded as infinite -- the linear behaviour remains unchanged. It can also be thought of as a limiting case of subcritical growth, in which the time it takes for equilibrium to be reached is infinitely long. It is thus clear that this regime of growth is by its nature only a border case between two more robust dynamical settings, so that trying to find an independent correspondence between it and a real life situation must be even tougher than in the supercritical case. On the other hand, from the point of view of dynamical systems, this kind of transition phenomenon might be of interest. 

\section{Conclusion \label{sec:conc}}

In summary, we have seen that an interesting diversity of dynamical patterns appear in our model, with higher or lower degrees of connection to ecological reality. The most important message is that all of them emerge from a single, biologically reasonable set of individual behaviour rules, in which one attempts to model the elementary features of food-limitation-induced intraspecific competition. Moreover, the transition among those different growth patterns is ruled, aside from initial conditions, by only three parameters -- all of straightforward biological interpretation: birth rate (the inverse of average generation time), available food per unit time (a measure of the environment's resource abundance) and the lifespan of a newborn in scarcity of food. In particular, in the ecologically appealing subcritical case, we have derived an expression for the environment carrying capacity as a function of those parameters (Eq. \ref{eq:eq_size}). All of this may indicate that a rich, useful, novel class of population growth models may be built upon the seminal ideas here addressed, by introducing punctual modifications in the behaviour rules that we have defined. Such adaptations might account for phenomena not captured by our model, such as extinction, inheritance, and even evolution. 

\section*{Acknowledgements}

L.R.T.N. thanks his fellow and friend Solano E. S. Fel\'icio for important discussions carried out along the development of the research here addressed. L.R.T.N.'s work has been supported by the S\~ao Paulo Research Foundation (FAPESP/Brazil) under grant no. 2018/13032-1.   


\appendix  

\section{ \label{sec:trans_ex}Transition expectations}

Taking the expected value in both sides of Eqs. \eqref{eq:class_std_dynamics}, conditioned to $\mathbf{X}_t = \mathbf{x}$, we obtain

    \begin{eqnarray}
        \label{eq:expectation_raw}
	    \ex{ X^i_{t+1} } = \begin{cases}
            x^{i+1} + \ex{ Q^i_{t}} - \ex{ Q^{i+1}_{t}}, & i <\omega; \\
            \ex{ Q^i_{t} }  + \ex{ R_{t} }, & i = \omega. \\
        \end{cases}
    \end{eqnarray}

As we are concerned only with competition regime, each $Q^i_t$ is hypergeometrically distributed with parameters $n, \alpha, x^i$, whence $\ex{Q^i_t} = (\alpha/n)x^i$. In its turn, $R_t$ has binomial distribution once the $Q^i_t$`s have been realized, say $Q^i_t = q^i$; for, $x^1 - q^1$ individuals will have died, the remaining $n - (x^1 - q^1)$ being able to generate a newborn with probability $\rho$. Thus,

    \begin{eqnarray}
        \label{eq:ex_Rt_cond}
        \ex{R_t \mid Q^1_t = q^1} = \rho \left[ n - (x^1 - q^1) \right]
    \end{eqnarray}
and, by the Total Probability Theorem,

    \begin{eqnarray}
        \label{eq:ex_Rt_tot}
        \ex{R_t} = \sum_{q^1} \ex{R_t \mid Q^1_t = q^1}\mathbb{P}_{\mathbf{x}}\left( Q^1_t=q^1\right)
    \end{eqnarray}

Using Eq. \eqref{eq:ex_Rt_cond} in Eq. \eqref{eq:ex_Rt_tot} and exploring the linearity of $\ex{R_t \mid Q^1_t=q^1}$ will give raise to two terms, one being the mean of $Q^1_t$ and the other, simply its normalization. Taking this result into Eq. \eqref{eq:expectation_raw} immediately yields Eq. \eqref{eq:expectation}. 

\section{\label{sec:critical}Analytic solution for critical growth}

Motivated by numerical results, we try in Eqs. \ref{eq:deterministic} a solution in which each $y^i_t$ grows linearly with time, with the restriction that all the $\omega$ angular coefficients be equal. That is, we try $y^i_t = at + b^i$; thus, $n_t = \omega a t + b, b \equiv \sum_i b^i$. Taking the form of $y^i_t$ into the first of Eqs. \eqref{eq:deterministic} gives 

    \begin{eqnarray}
    \label{eq:crit:1}
        b^{i+1}-b^i = \frac{a}{1- \alpha/n_t}.
    \end{eqnarray}

In an exact sense, Eq. \eqref{eq:crit:1} implies that no such solution may exist, for the left-hand side is a constant, whereas the right-hand depends upon $t$ (we have seen in \S \ref{sec:equilibrium} that $n_t$ cannot be stationary in critical regime). Nevertheless, since $n_t$ grows indefinitely with time in our \textit{ansatz}, we may assume that our solution only holds asymptotically for large values of $t$, so one may approximate $1 - \alpha/n_t \approx 1$ whence Eq. \eqref{eq:crit:1} reads $b^{i+1} = a + b^i, i = 0, ..., \omega - 1$. We thus write $b^i = b^0 + ai$ and $b = \omega b^0 + a\omega (\omega + 1)/2$.

Using the aimed expressions for $y^i_t$ and $n_t$ along with the obtained form of $b^i$ and $b$ in the second of Eqs. \eqref{eq:deterministic}, also imposing the critical relation $\rho(\omega -1) = 1$, one obtains

    \begin{eqnarray}
        a = \frac{2\rho \alpha}{\omega}\left[ \frac{\omega at + \omega b^0 + a\left( 1 + \omega/\rho\right)}{\omega at + \omega b^0 + a \omega ( \omega + 1 )/2 } \right].
    \end{eqnarray}

Again, the only way of satisfying this condition (unless we were to assume very specific values of $\rho$ and $\omega$) is if we take large values of $t$, such that the expression in brackets approaches $1$ and we have

    \begin{eqnarray}
        a = \frac{2\rho \alpha}{\omega}.
    \end{eqnarray}

\bibliographystyle{unsrt} 
\bibliography{bibliography}

\providecommand{\noopsort}[1]{}\providecommand{\singleletter}[1]{#1}%
\begin{thebibliography}{10}

\bibitem{nowak-2006}
M.~Nowak.
\newblock {\em Evolutionary Dynamics: Exploring the Equations of Life}.
\newblock Belknap Press, 2006.

\bibitem{strogatz-2015}
S.~H. Strogatz.
\newblock {\em Nonlinear Dynamics and Chaos: With Applications to Physics,
  Biology, Chemistry, and Engineering}.
\newblock Westview Press, 2nd edition, 2015.

\bibitem{hirsch-etal-2012}
R.~L.~Devaney M.~W.~Hirsch, S.~Smale.
\newblock {\em Differential Equations, Dynamical Systems, and an Introduction
  to Chaos}.
\newblock Academic Press, 3rd edition, 2012.

\bibitem{may-1976}
R.~May.
\newblock {\em Nature}, 261:459–467, 1976.

\bibitem{wade-2016}
M.~J. Wade, J.~Harmand, B.~Benyahia, T.~Bouchez, S.~Chaillou, B.~Cloez, J.-J.
  Godon, B.~M. Boudjemaa, A.~Rapaport, T.~Sari ad~R.~Arditi, and C.~Lobry.
\newblock {\em Ecological Modelling}, 321:64--74, 2016.

\bibitem{shoener-2019}
B.~D. Shoener, S.~M. Schramm, F.~Béline, O.~Bernard, C.~Martínez, B.~G.
  Plósz, S.~Snowling, J.-P. Steyer, B.~Valverde-Pérez, D.~Wágner, and J.~S.
  Guest.
\newblock {\em Water Research X}, 2:100024, 2019.

\bibitem{mckane-2004}
A.~J. McKane and T.~J. Newman.
\newblock {\em Physical Review E}, 70:041902, 2004.

\bibitem{deangelis-2005}
D.~L. DeAngelis and W.~M. Mooij.
\newblock {\em Annual Review of Ecology, Evolution, and Systematics},
  36:147–162, 2005.

\bibitem{law-2003}
Richard Law, David~J. Murrell, and Ulf Dieckmann.
\newblock {\em Ecology}, 84(1):252--262, 2003.

\bibitem{natalie-2017}
N.~Tkachenko, J.~D. Weissmann, W.~P. Petersen, G.~Lake, C.~P.~E. Zollikofer,
  and S.~Callegari.
\newblock {\em PLoS ONE}, 12(4):e0176101, 2017.

\bibitem{nasell-2001}
I.~Nasell.
\newblock {\em Journal of Theoretical Biology}, 211:11--27, 2001.

\bibitem{otso-2010}
O.~Ovaskainen and B.~Meerson.
\newblock {\em Trends in Ecology \& Evolution}, 25:643–652, 2010.

\bibitem{maas-1978}
A.~Cunningham and P.~Maas.
\newblock {\em Journal of General Microbiology}, 104:227--231, 1978.

\bibitem{tang-1997}
B.~Tang, A.~Sitomer, and T.~Jackson.
\newblock {\em Journal of Mathematical Biology}, 35:453--479, 1997.

\bibitem{krebs-2014}
C.~J. Krebs.
\newblock {\em Ecology: The Experimental Analysis of Distribution and
  Abundance}.
\newblock Pearson, 6th edition, 2014.

\bibitem{pearl-1920}
R.~Pearl and L.~J. Reed.
\newblock {\em PNAS}, 6:275--288, 1920.

\end{thebibliography}

\end{document}